\documentclass[10pt,twoside]{classe-cmb}
\usepackage{amssymb,amsbsy,amsmath,amsfonts,amssymb,amscd}
\usepackage{latexsym,euscript,exscale,epic,eepic,epsfig}
\usepackage[english,francais]{babel}
\usepackage{times}
\def\vec#1{\ensuremath{\mathbf{#1}}}

\newtheorem{e-proposition}[theorem]{Proposition}

\newtheorem{e-definition}[theorem]{Definition\rm}

\ComParit{S1296-2147}
\PIT{FLA}
\PXHY{????}
\Add{?}
\Volume{0}
\Year{2003}
\FirstPage{1}
\LastPage{??}
\AuteurCourant{J.-Ch. Hamilton}
\TitreCourant{CMB map-making and power spectrum estimation} 
\Journal
\Rubrique{Rubrique}{Heading}
\SousRubrique{Sous-rubrique}{Sub-Heading}
\PresentePar{Jean-Christophe}{HAMILTON}
\Recu{blank}{}{}
\begin{document}
\selectlanguage{english}
\TitleOfDossier{The Cosmic Microwave Background}
\title{%
CMB map-making and power spectrum estimation
}
\author{%
Jean-Christophe HAMILTON~$^{\text{a}}$,\ \
}
\address{%
\begin{itemize}\labelsep=2mm\leftskip=-5mm
\item[$^{\text{a}}$]
LPNHE (CNRS-IN2P3), Paris VI \& VII - 4, Place Jussieu, 75252 Paris Cedex 05 - France\\
\item[$^{\text{a}}$]
LPSC (CNRS-IN2P3), 53, Avenue des Martyrs, 38026 Grenoble Cedex - France\\
\item[$^{\text{a}}$]
PCC (CNRS-IN2P3), 11, Place Marcelon Berthelot, 75231 Paris Cedex 05 - France\\
E-mail: hamilton@in2p3.fr
\end{itemize}
}
\maketitle
\thispagestyle{empty}
\begin{Abstract}{%
CMB data analysis is in general done through two main steps :
map-making of the time data streams and power spectrum extraction from
the maps. The latter basically consists in the separation between the
variance of the CMB and that of the noise in the map. Noise must
therefore be deeply understood so that the estimation of CMB variance
(the power spectrum) is unbiased. I present in this article general
techniques to make maps from time streams and to extract the power
spectrum from them. We will see that exact, maximum likelihood
solutions are in general too slow and hard to deal with to be used in
modern experiments such as Archeops and should be replaced by
approximate, iterative or Monte-Carlo approaches that lead to similar
precision.}
\end{Abstract}

\par\medskip\centerline{\rule{2cm}{0.2mm}}\medskip
\setcounter{section}{0}
\selectlanguage{english}
\section{Introduction}
The cosmological information contained in the Cosmic Microwave
Bacground (CMB) anisotropies is encoded in the angular size
distribution of the anisotropies, hence in the angular power spectrum
and noted $C_\ell$.  It is of great importance to be able to compute
the $C_\ell$ spectrum in an unbiased way. The simplest procedure to
obtain the power spectrum is to first construct a map of the CMBA from
the data timelines giving the measured temperature in one direction of
the sky following a given scanning strategy on the sky, this is known
as the map-making process ; then extract the $C_\ell$ from this map,
this is the power spectrum extraction. Various effects usually present
in the CMB data make these two operations non trivial. The major
effect being related to the unavoidable presence of instrumental and
photon noise. Noise in the timelines is correlated and appears as low
frequency drifts that are still present in the map. A good map-making
process minimizes these drifts, but in most cases, they are still
present in the map. They have to be accounted for in the power
spectrum estimation as the signal power spectrum is nothing but an
excess variance in the map at certain angular scales compared to the
variance expected from the noise. The CMBA power spectrum will
therefore be unbiased only if the noise properties are known
precisely.

This article presents the usual techniques that allow an unbiased
determination of both the CMBA maps and power spectrum. In
Sect.~\ref{datamodel} we will describe the data model and the data
statistical properties required for the techniques presented here to
be valid. Sections~\ref{mapmaking} and \ref{clspectrum} respectively
deal with map-making and power spectrum estimation techniques.

\section{Data model~\label{datamodel}}
The initial data are time ordered information (TOI) taken along the
scanning strategy pattern of the experiment. The detector measures the
temperature of the sky in a given direction through an instrumental
beam. This is equivalent to say that the underlying sky is convolved
with this instrumental beam and that the instrument measures the
temperature in a single direction of a $N_p$ pixellised convolved sky
noted $\vec{T}$. The $N_t$ elements TOI noted $\vec{d}$ may therefore
be modelled as:
\begin{equation}\label{defmod}
\vec{d}=A\cdot\vec{T}+\vec{n}
\end{equation}
The pointing matrix $A$ relates each time sample to the corresponding
pixel in the sky. $A$ is a $N_t\times N_p$ matrix that contains a
single 1 in each line as each time sample is sensitive to only one
pixel is the convolved sky\footnote{Different forms for $A$ can
however be used in case of differential measurements or more complex
scanning strategies.}. The noise TOI $\vec{n}$ in general has a non
diagonal covariance matrix $N$ given by\footnote{the symbols $\left<~\right>$ mean that we take the ensemble
average over an infinite number of realisations.}:
\begin{equation}
N=\left<\vec{n}\cdot\vec{n}^t\right>
\end{equation}
The most important property of the noise, that will be used widely
later is that it {\em has to be} Gaussian and piece-wise
stationary. Both assumptions are crucial as they allow major
simplifications of the map-making and power spectrum estimation
problems, namely Gaussianity means that all the statistical
information on the noise is contained in its covariance matrix and
stationarity means that all information is also contained in its
Fourier power spectrum leading to major simplifications of the
covariance matrix : the noise depends only on the time difference
between two samples and $N$ is therefore a Toeplitz matrix
$N_{ij}=N_{\left|i-j\right|}$ completely defined by its first line and
is very close to be circulant\footnote{Saying that the matrix is
circulant is an additionnal hypothesis, but a very good approximation
for large matrices.}. Such a matrix is diagonal in Fourier space.  Its
first line is given by the autocorrelation function of the noise, that
is the inverse Fourier transform of its Fourier power spectrum
($\star$ is the convolution operator)\footnote{$\cal{F}$ denotes the Fourier transform (in practice, a FFT
algorithm is used).}:
\begin{equation}
N_{i0}=\left<\vec{n}\star\vec{n}\right>\equiv \left<{\cal{F}}^{-1}\left[\left|{\cal{F}}(\vec{n})\right|^2\right]\right>
\end{equation}

\section{Map-making techniques\label{mapmaking}}
The map-making problem is that of finding the best estimate
$\hat{\vec{T}}$ of $\vec{T}$ from Eq.~\ref{defmod} given $\vec{d}$ and
$A$. The noise $\vec{n}$ is of course unknown. We will address the two
main approaches to this problem, the first being the simplest one and
the second one being the optimal one. An excellent detailed review on
map-making techniques for the experts is~\cite{stompor2002}.

\subsection{Simplest map-making : coaddition}
The simplest map-making that one can think about is to neglect the
effects of the correlation of the noise. One can just average the data
falling into each pixel without weighting them. This procedure is
optimal (it maximises the likelihood) if the noise in each data sample
is independant, that is, if the noise is white. In a matrix notation,
this simple map-making can be written:
\begin{equation}
\hat{\vec{T}}=\left[A^t\cdot A\right]^{-1}\cdot A^t \cdot \vec{d}
\end{equation}
where the operator $A^t$ just projects the data into the correct pixel
and $\left[ A^t\cdot A\right]$ counts the sample falling into each
pixel. This simple map-making has the great advantage of the
simplicity. It is fast ($\propto N_t$) and robust. 

However, in the case of realistic correlated noise, the low frequency
drifts in the timelines induce stripes in the maps along the scans of
the experiment. These stripes are often much larger than the CMBA
signal that is searched for and therefore should be avoided. Various
destriping techniques have been proposed to avoid these stripes. A
method exploiting the redundancies of the Planck mission\footnote{\tt
http://astro.estec.esa.nl/Planck/} scanning strategy has been proposed
by~\cite{delabrouille98} and extended to polarisation
by~\cite{revenu2000}. This kind of method aims at suppressing the low
frequency signal by requiring that all measurements done in the same
direction at different instant coincide to a same temperature
signal. Another method has recently been proposed for the
Archeops\footnote{\tt http://www.archeops.org/} data analysis and
estimates the low frequency drifts by minimizing the cross-scan
variations in the map due to the drifts~\cite{bourrachot2003}.  The
simplest method for removing the low frequency drifts before applying
simple map-making is certainly to filter the timelines so that the
resulting timeline has almost white noise. The filtering can consist
in prewhitening the noise or directly setting to zero contaminated
frequencies. The computing time (CPU) scaling of the filtering +
coaddition process is modest and dominated by filtering ($\propto
N_t\log N_t$). This method however removes also part of the signal on
the sky and induces ringing around bright sources which has
to be accounted for in later processes.

\subsection{Optimal map-making}
The most general solution to the map-making problem is obtained by
maximizing the likelihood of the data given a noise
model~\cite{wright96,tegmark97}. As the noise is Gaussian, its
probability distribution is given by the $N_t$ dimensionnal Gaussian:
\begin{equation}
P(\vec{n})=\frac{1}{\left|(2\pi)^{N_t}
N\right|^{1/2}}\exp\left[-\frac{1}{2}\vec{n}^t\cdot
N^{-1}\cdot\vec{n}\right]
\end{equation}
Assuming no prior on the sky temperature, one gets from
Eq.~\ref{defmod} the probability of the sky given the data:
\begin{equation}
P(\vec{T}|\vec{d})\propto P(\vec{d}|\vec{T})\propto\frac{1}{\left|(2\pi)^{N_t}
N\right|^{1/2}}\exp\left[-\frac{1}{2}\left(\vec{d}-A\cdot\vec{T}\right)^t\cdot
N^{-1}\cdot\left(\vec{d}-A\cdot\vec{T}\right)\right]
\end{equation}
Maximizing this probability with respect to the map leads to solving
the linear equation:
\begin{equation}\label{lineq}
A^t\cdot N^{-1}\cdot A \cdot \vec{T}=A^{t}\cdot N^{-1}\cdot\vec{d}
\end{equation}
with solution\footnote{One can remark here that simple map-making is
equivalent to optimal map-making if the noise covariance matrix is
diagonal, which is consistent to what was said before.}:
\begin{equation}
\hat{\vec{T}}=\left(A^t\cdot N^{-1}\cdot A\right)^{-1}\cdot A^{t}\cdot N^{-1}\cdot\vec{d}
\end{equation}
One therefore just has to apply this linear operator to the data
timeline to get the best estimator of $\vec{T}$, note that
$\hat{\vec{T}}$ is also the minimum variance estimate of the map. The
covariance matrix of the map is:
\begin{equation}
{\cal N}=\left(A^t\cdot N^{-1}\cdot A\right)^{-1}
\end{equation}

Problems arise when trying to implement this simple procedure, the
timeline data and the maps are in general very large : the typical
dimensions of the problem are $N_t\simeq 6\times 10^7$ and $N_p\simeq
10^5$ for Archeops.

The maximum likelihood solution requires both $N^{-1}$ and
$\left(A^t\cdot N^{-1}\cdot A\right)^{-1}$ which are not easy to
determine. Two approaches can be used at this point: one can try to
make a brute force inversion of the problem, relying on huge parallel
computers or one can try to iteratively approach the solution, hoping
that convergence can be reached within reasonnable time.

\subsection{Brute force inversion}
The brute force optimal map-making parallel implementation is freely
available as the MADCAP \cite{madcap} package. It is a general
software designed to produce an optimal map for any experiment by
solving directly Eq.~\ref{lineq}. The use of this package requires the
access to large parallel computers.

The only assumption that is done in MADCAP map-making is that the
inverse time-time noise covariance matrix can be obtained directly
without inversion from the noise Fourier power spectrum:
\begin{equation}
N^{-1}_{i0}\simeq \left<{\cal{F}}^{-1}\left[\frac{1}{\left|{\cal{F}}(\vec{n})\right|^2}\right]\right>
\end{equation}
This assumption is not perfectly correct on the edges of the matrix
but leads to a good estimate of the inverse time covariance
matrix for the sizes we deal with. This allows this step
to scale as $N_t\log N_t$ operations rather than the $N_t^2$ required
by a Toeplitz matrix inversion. In most cases, the time correlation
$N_\tau$ length is less than the whole timestream $N_t$ so that $N$ is
band-diagonal. For Archeops, we have $N_\tau\simeq 10^4$.

The next step is to compute the inverse pixel noise covariance matrix
${\cal N}^{-1}=\left(A^t\cdot N^{-1}\cdot A\right)$ and the noise weighted map
$A^{t}\cdot N^{-1}\cdot\vec{d}$, both operations scale as $N_t
\times N_\tau$ when exploiting the structure of $A$ and $N$. The 
last step is to invert ${\cal N}^{-1}$ and multiply it by $A^{t}\cdot
N^{-1}\cdot\vec{d}$ to get the optimal map. Unfortunately, ${\cal
N}^{-1}$ has no particular structure that can be exploited and
this last step scales as a usual matrix inversion $\propto N_p^3$ and
largely dominates the CPU required by MADCAP for the usual
large datasets ({\em eg.} Archeops).

We can remark here that MADCAP provides the map covariance matrix
${\cal N}$ for free as a byproduct. This matrix is crucial for
estimating the power spectrum as will be seen in
section~\ref{clspectrum}.

\subsection{Iterative solutions}
The other possibility is to solve Eq.~\ref{lineq} through an iterative
process such as the Jacobi iterator, or more efficiently a
conjugate-gradient~\cite{numrec}. Both converge to the
maximum likelihood solution.

The use of the Jacobi iterator for solving for the maximum likelihood
map in CMB analysis was first proposed by~\cite{prunet}. The basic
algorithm is the following. We have to solve the following linear
system (see Eq.~\ref{lineq}):
\begin{equation}
\Gamma\cdot\vec{x}=\vec{y}
\end{equation}
The Jacobi iterator starts with an approximation
$\Lambda_0$ of $\Gamma^{-1}$ and iterates to improve the residuals $R$:
\begin{equation}
\Lambda_0\cdot \Gamma=I-R
\end{equation}
In order to converge, the algorithm requires the first approximation
to be good enough so that the eigenvalues of $R$ are all smaller than
1 (a good estimate in general is
$\Lambda_0=\left[\mathrm{diag}\Gamma\right]^{-1}$). We can therefore
expand:
\begin{equation}
\Gamma^{-1}=\left(I-R\right)^{-1}\cdot\Lambda_0
=\left(I+R+R^2+\cdots\right)\cdot\Lambda_0
\end{equation}
Lets us define
$\Lambda_n=\left(I+R+R^2+\cdots+R^n\right)\cdot\Lambda_0$ so that
$\Gamma^{-1}=\lim_{n\rightarrow\infty}\Lambda_n$. We have the relationship
$\Lambda_{j+1}=\Lambda_0+R\cdot\Lambda_j$. If we define
$\vec{x}_j=\Lambda_j\cdot \vec{y}$, it is straightforward to show
that:
\begin{equation}\label{jacobieq}
\vec{x}_{j+1}-\vec{x}_j=\Lambda_0\cdot\left(\vec{y}-\Gamma\cdot\vec{x}_j\right)
\end{equation}
which defines the Jacobi iterator. When going back to the usual CMB
notation for maps and timelines, one gets:
\begin{equation}
\vec{T}_{j+1}-\vec{T}_j=\left[\mathrm{diag}\left(A^t\cdot N^{-1}\cdot A\right)\right]^{-1}\cdot A^t\cdot N^{-1}\cdot \left(\vec{d}-A\cdot\vec{T}_j\right)
\end{equation}
which looks rather complicated but is in fact very simple to
implement: the operation $A\cdot\vec{T}_j$ just consists in reading
the map at iteration $j$ with the scanning strategy ($\propto N_t$),
and the matrix $\mathrm{diag}\left(A^t\cdot N^{-1}\cdot A\right)$ is
just the white noise level variance divided by the number of hits in
each pixel. It is diagonal and therefore does not require proper
inversion. The only tricky part here is the multiplication
$N^{-1}\cdot \left(\vec{d}-A\cdot\vec{T}_j\right)$ given the fact that
$N^{-1}$ is unknown. As the noise is stationary, $N$ is Toeplitz and
circulant\footnote{again, it is not exactely circulant but it is an
excellent approximation as the matrix is large}, the multiplication by
$N^{-1}$ can be done in Fourier space directly through:
\begin{equation}
N^{-1}\cdot\vec{x}\simeq {\cal F}^{-1}\left[\frac{{\cal F}(\vec{x})}{\left|{\cal F}(\vec{x})\right|^2}\right]
\end{equation}
which requires $N_t\log N_t$ operations. Finally, each iteration is
largely dominated by the latter so that the final CPU time scales like
$N_{it}\times N_t\log N_t$ where $N_{it}$ is the number of iterations. 

Unfortunately the convergence of such an iterator is very slow and
makes it rather unefficient as it is. A significant improvement was
proposed by \cite{mapcumba} in the publicly available software
MAPCUMBA. They noted that the convergence was actually very fast on
small scales (compared to the pixel) but that the larger scales were
converging slowly. They proposed a multigrid method where the pixel
size changes at each iteration so that the global convergence is
greatly accelerated (see Fig.~7 of \cite{mapcumba}), making this
iterative map-making really efficient. A conjugate gradient solver
instead of the Jacobi iterator is implemented in the software
Mirage~\cite{mirage} and accelerates again the convergence
significantly. A new version of MAPCUMBA also uses a conjugate
gradient solver, as well as MADmap~\cite{cantalupo}.

If obtaining an optimal map is now quite an easy task using an
iterative implementation (the presence of strong sources, such as the
galactic signal however complicates this simple picture), they do not
provide the map noise covariance matrix ${\cal N}=\left(A^t\cdot
N^{-1}\cdot A\right)^{-1}$ which is of great importance when computing
the CMB power spectrum in the map in order to be able to make the
difference between noise fluctuations and real signal
fluctuations. The only way to obtain this covariance matrix using
these iterative methods is through large Monte-Carlo simulation that
would reduce the advantage of iterative map-making compared to
brute-force map-making.

\subsection{map-making comparisons}

\begin{figure}[!h]
\centering \resizebox{12cm}{!}
{\includegraphics[clip,width=19cm]{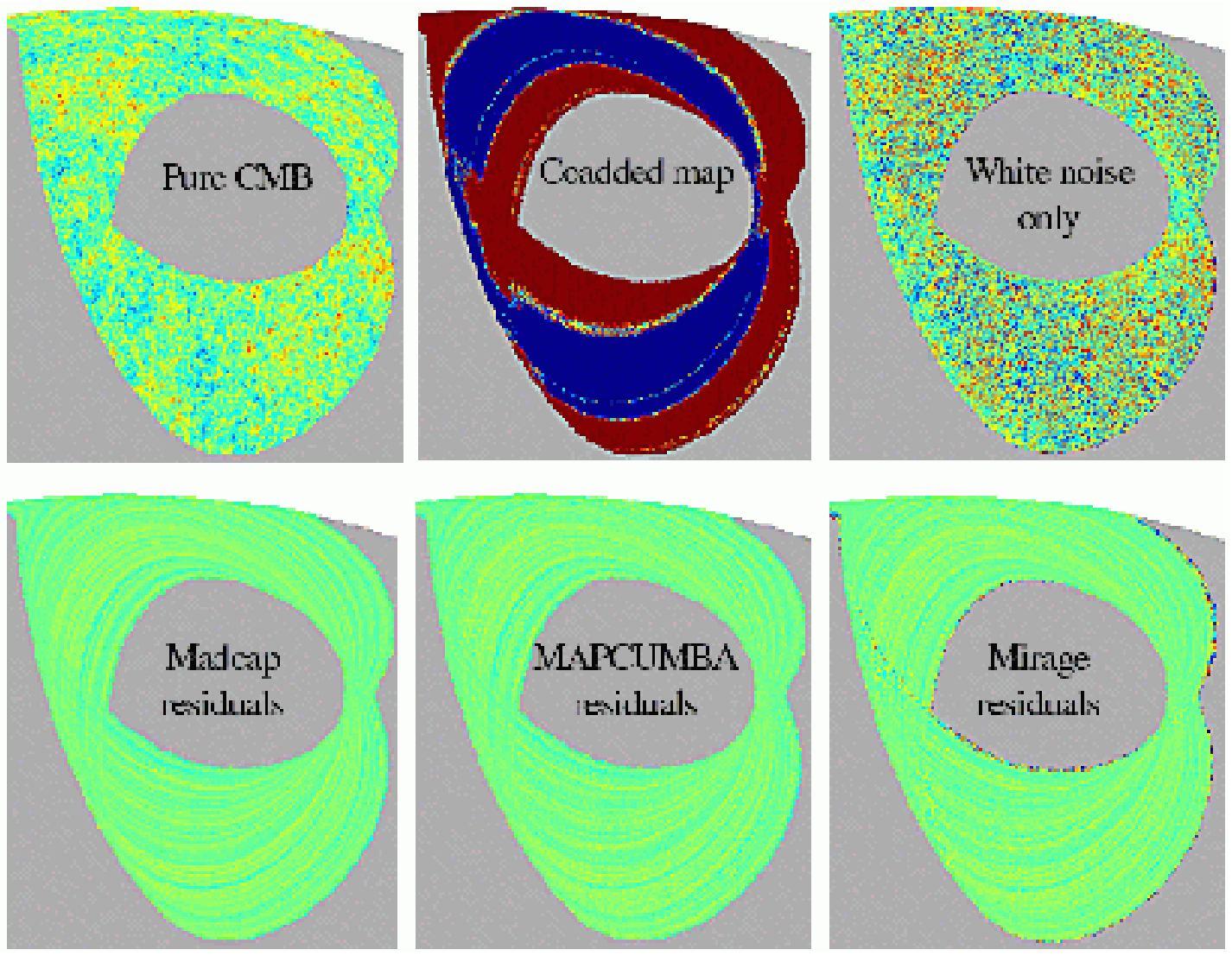}\includegraphics[clip,width=22cm]{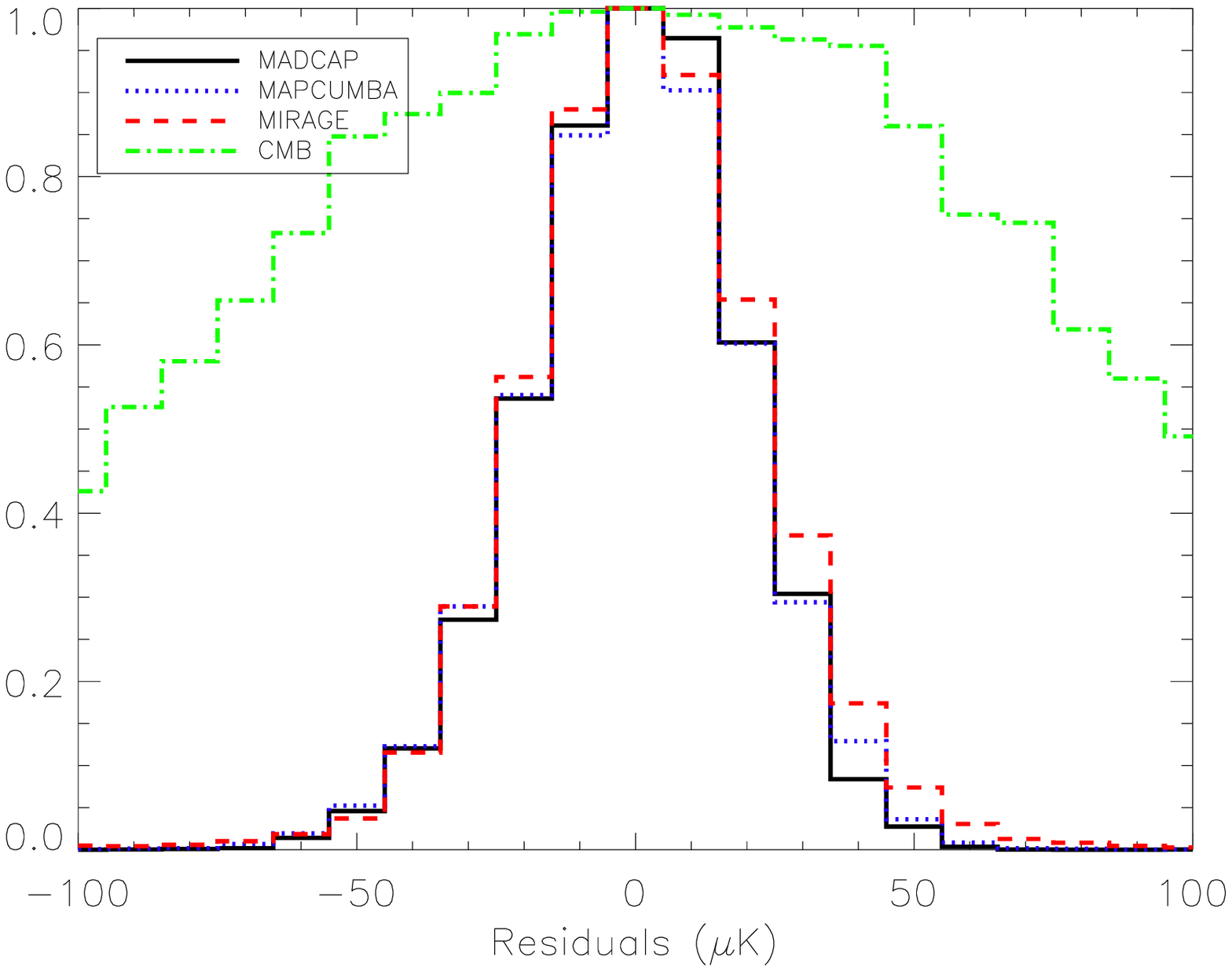}}
\caption{\small The six maps on the right show a comparison of results from 
different map-making implementations on the same simulation (typical
of Archeops data). All maps are in Healpix
pixellisation~\cite{healpix} and have the same color scale ranging
from -500 $\mu\mathrm{K}$ to 500 $\mu\mathrm{K}$ from dark blue to
dark red, green corresponds to zero. The histograms of the residuals
is shown on the right and is more than three times smaller than the
actual CMB fluctuation.}
\label{fig:cmp}
\end{figure}

The precision of the MADCAP, MAPCUMBA and Mirage implementations are
shown in Fig~\ref{fig:cmp} with the same CMB and noise simulation
based on Archeops realistic conditions. The three resulting maps were
kindly provided by~\cite{filliatre}.  The six maps on the left are
respectively from top left to bottom right : initial CMB fluctuation,
coaddition of the timeline without filtering, coaddition of the
timeline with white noise only ({\it ie} the true optimal map that has
to be reconstructed), MADCAP residual map (difference between MADCAP
reconstructed map and the white noise map), MAPCUMBA residual map and
Mirage residual map. All maps are shown with the same color scale. The
first remark that can be done is that the stripes are indeed a real
problem and that straight coaddition is not to be performed. The three
different optimal map-making codes give very similar results,
especially MADCAP and MAPCUMBA. In all cases, as can be also seen in
the right panel of Fig.~\ref{fig:cmp}, the residuals are much smaller
than the CMB fluctuations that are searched for. The three map-making
implementation can therefore be considered are unbiased\footnote{Let's
note that the noise model that was used for MADCAP is the true one,
not an estimation. This makes however little difference.}.

Finally one can summarize the comparison as following, iterative and
brute-force optimal map-making give very similar results as far the
optimal map is concerned. The brute force inversion provides the map
noise covariance matrix for free which is a major point as will be
seen in next section. The computer requirements are however much
larger than for iterative map-making. The latter should therefore be
used when the power spectrum estimation can be carried out without the
knowledge of the map noise covariance matrix, in general using a
Monte-Carlo technique (see next section). In this case, one should
seriously consider the filtering + coaddition map-making that is by
far the fastest but removes part of the signal. This is however
accounted for (see section~\ref{frequentist}) also using a Monte-Carlo
technique.

\section{Power spectrum estimation techniques\label{clspectrum}}
We know want to compute the power spectrum $C_\ell$ of the map
$\vec{T}$ whose noise covariance matrix ${\cal N}$ might be known or
not depending on the method that was used before to produce the
map. The map is composed of noise and signal (from now on $\vec{n}$ is
the noise on the map pixels):
\begin{equation}
\vec{T}=\vec{s}+\vec{n}
\end{equation}
The signal in pixel $p$ can be expanded on the $Y_{\ell
m}\left(\theta_p,\phi_p\right)$ spherical harmonics basis :
\begin{equation}\label{sph_harm}
s_p=\sum_{\ell=0}^{\infty}\sum_{m=-\ell}^\ell a_{\ell m} B_\ell Y_{\ell m}\left(\theta_p,\phi_p\right)
\end{equation}
where $B_\ell$ stands for the beam\footnote{It is the Legendre
transform of the instrumental beam under the assumption that it is
symmetric.}.  If the CMBA are Gaussian, the variance of the $a_{\ell
m}$, called the angular power spectrum and denoted $C_\ell$ contains
all the cosmological information :
\begin{equation}
\left<a_{\ell m}a_{\ell^\prime
m^\prime}^\star\right>=C_\ell \delta_{\ell \ell^\prime}\delta_{m
m^\prime}
\end{equation}
The map covariance matrix (assuming no correlation between signal and
noise) is:
\begin{eqnarray}
M&=&\left<\vec{T}\cdot\vec{T}^t\right>=\left<\vec{s}\cdot\vec{s}^t\right>+\left<\vec{n}\cdot\vec{n}^t\right>\\
&=&S+N
\end{eqnarray}
and the signal part is related to the $C_\ell$:
\begin{equation}
S_{pp^\prime}=\left<s_p s_{p^\prime}\right> =\sum_\ell \frac{2\ell+1}{4\pi}C_\ell B_\ell^2 P_\ell(\chi_{pp^\prime}) \label{sdef}
\end{equation}
where
$\chi_{pp^\prime}=\cos\left(\vec{u}_p\cdot\vec{u}_{p^\prime}\right)$,
$\vec{u}_p$ being the unit vector towards pixel $p$ and $P_\ell$ are
the Legendre polynomials.

One therefore has a direct relation between the map and noise covariance matrices and the angular power spectrum:
\begin{equation}\label{powspec_relation}
M=N+\sum_\ell \frac{2\ell+1}{4\pi}C_\ell B_\ell^2 P_\ell(\chi_{pp^\prime})
\end{equation}
The power spectrum estimation consists in estimating $C_\ell$ from
$\vec{T}$ and $N$ (that can be unknown) using this relation.

\subsection{Maximum likelihood solution}
Full details concerning this can be found in~\cite{bjk,tegmark}. As
for the map-making problem, the maximum likelihood solution proceeds
by writing the probability for the map given its covariance matrix
assuming Gaussian statistics\footnote{The trace appears from
$\left|M\right|^{-1}=\exp\left[-\mathrm{Tr}\left(\ln
M\right)\right]$ as the trace is invariant.}:
\begin{equation}
P(C_\ell|\vec{T})\propto P(\vec{T}|C_\ell)=\left(2\pi\right)^{-N_{p}/2}
\exp \left[ -\frac{1}{2}\left[\left(\vec{T}^t\cdot M^{-1}\cdot \vec{T}\right)
+\mathrm{Tr}\left(\ln M\right) \right]\right]
\end{equation}
and we therefore want to maximize the likelihood function through
$\frac{\partial L}{\partial C_\ell}=0$:
\begin{equation}\label{likelihood}
L(C_\ell)= -\frac{1}{2}\left[\left(\vec{T}^t\cdot M^{-1}\cdot \vec{T}\right)
+\mathrm{Tr}\left(\ln M\right)\right]
\end{equation}
Tedious calculations lead to the solution:
\begin{equation}\label{eqcl}
C_\ell=\sum_{\ell^\prime}F^{-1}_{\ell\ell^\prime}\times
\mathrm{Tr}\left[\left(\vec{T}\cdot\vec{T}^t-N\right)
\cdot M^{-1}\cdot \frac{\partial S}{\partial C_\ell}\cdot M^{-1}\right]
\end{equation}
where $F$ is the Fisher matrix:
\begin{equation}
F_{\ell\ell^\prime}=
\mathrm{Tr}\left[\frac{\partial S}{\partial C_\ell}\cdot M^{-1}
\cdot \frac{\partial S}{\partial C_{\ell^\prime}}\cdot M^{-1}\right]
\end{equation}
Eq.~(\ref{eqcl}) let $C_\ell$ appear in both sides (in $M$) in an
uncomfortable way and therefore cannot be solved simply. The method
usually used~\cite{bjk,madcap} is the Newton-Raphson iterative scheme:
One starts from an initial guess for the binned power
spectrum\footnote{Binned power spectrum means that we do not consider
one single mode $\ell$ but a bin in $\ell$ as we do not have access in
general to all modes due to incomplete sky coverage.} $\vec{C}^{(0)}$ and
iterates until convergence following:
\begin{equation}
\vec{C}^{(i+1)}=\vec{C}^{(i)}+\delta\vec{C}
\end{equation}
with:
\begin{equation}\label{itcl}
\delta\vec{C}=-\left[ \left.  \frac{\partial^2 L}{\partial \vec{C}^2}
       \right| _{\vec{C}=\vec{C}_i}
\right]^{-1} \cdot
\left. \frac{\partial L}{\partial \vec{C}}
\right|_{\vec{C}=\vec{C}_i}
\end{equation}
the likelihood $L$ being that of Eq.~\ref{likelihood}. Convergence is
usually reached after a few iterations. The explicit form of the
derivatives of Eq.~\ref{itcl} is:
\begin{eqnarray}
\frac{\partial L}{\partial C_b}&=&
\frac{1}{2}\left(\vec{m}^T\cdot M^{-1}\cdot
\frac{\partial S}{\partial C_b}\cdot M^{-1}\cdot\vec{m}-
\mathrm{Tr}\left[M^{-1}\cdot\frac{\partial S}{\partial C_b}\right]\right)\\
\frac{\partial^2 L}{\partial C_b\partial C_{b^\prime}}&=&
-\vec{m}^T\cdot M^{-1}\cdot\frac{\partial S}{\partial C_b}\cdot M^{-1}\cdot
\frac{\partial S}{\partial C_{b^\prime}}\cdot M^{-1}\cdot\vec{m}
+\frac{1}{2}\mathrm{Tr}\left[M^{-1}\cdot
\frac{\partial S}{\partial C_b}\cdot M^{-1}\cdot
\frac{\partial S}{\partial C_{b^\prime}}\right]
\end{eqnarray}
where the index $b$ denotes the bin number.

Each iteration will then require a large number of large matrix
operations forcing such an algorithm to be implemented on large memory
parallel supercomputers. MADCAP~\cite{madcap} is the common
implementation of this algorithm and scales as
$2(N_b+\frac{2}{3})N_p^3$ operations per iteration. The CPU/RAM/Disk
problem is therefore even cruder for the power spectrum than for the
map-making. This algorithm leads to the optimal solution accounting
correctly for the noise covariance matrix and additionnaly provides
the likelihood shape for each bin through the various iterations
allowing to a direct estimate of the error bars.

\subsection{Frequentist approaches\label{frequentist}}
An alternative approach to power spectrum estimation is to compute the
so called pseudo power spectrum (harmonic transform of the map, noted
$\tilde{C_\ell}$) and to correct it so that it becomes a real power
spectrum. This approaches have been proposed and developped
in~\cite{hivon,spice}. The harmonic transform of the map differs from
the true $C_\ell$ in various ways (we follow the notations
from~\cite{hivon}): The observed sky is convolved by the beam and by
the transfer function of the experiment so that the observed power
spectrum is $B_\ell^2 F_\ell C_\ell$, where $B_\ell$ characterizes the
beam shape in harmonic space and $F_\ell$ the filtering done to the
data by the analysis process (that may also include electronic
filtering by the instrument itself).  The observed sky is in general
incomplete (at least because of a Galactic cut) leading to the fact
that the $C_\ell$ measured are not independant as they are convolved
in harmonic space by the window-function~\cite{white_srednicki}. We
therefore have access to $\sum_{\ell^\prime}
M_{\ell\ell^\prime}B_{\ell^\prime}^2 F_{\ell^\prime}C_{\ell^\prime}$
where $M_{\ell\ell^\prime}$ is the mode mixing matrix.  Finally, the
noise in the timelines projects on the sky and adds its contribution
$\tilde{N}_\ell$ to the sky angular power spectrum.  At the end, the
map angular power spectrum, the {\em pseudo-$C_\ell$} is related to
the true $C_\ell$ via: \begin{equation} \label{eqmaster}
\tilde{C}_\ell=\sum_{\ell^\prime} M_{\ell\ell^\prime}B_{\ell^\prime}^2
F_{\ell^\prime}C_{\ell^\prime} +\tilde{N}_\ell \end{equation}

The frequentist methods propose to invert Eq.~\ref{eqmaster} making an
extensive use of Monte-Carlo simulations (details can be found
in~\cite{hivon}):
\begin{itemize}
\item the pseudo power spectrum of the map $\tilde{C_\ell}$ is computed by 
transforming the map into spherical harmonics (generally using Healpix
pixellisation and the {\tt anafast} procedure available in the Healpix
package~\cite{healpix}).
\item The mode mixing matrix is computed analytically through:
\begin{equation}
M_{\ell_1
\ell_2}=\frac{2\ell_2+1}{4\pi}\sum_{\ell_3}\left(2\ell_3+1\right){\mathcal
W}_{\ell_3}\left(
\begin{array}{ccc}
\ell_1 & \ell_2 & \ell_3 \\
0      & 0      & 0
\end{array}\right)
\end{equation}
where ${\mathcal W}_\ell$ is the power spectrum of the window of the
experiment (in the simplest case 1 for the observed pixels and 0
elsewhere, but more complex weighting schemes may be used, as in
Archeops \cite{arch_cell} or WMAP~\cite{wmap_cell}). In the SpICE
approach~\cite{spice}, the $M_{\ell \ell^\prime}$ inversion in
harmonic space is replaced by a division in angular space which is
mathematically equivalent.
\item The beam transfer function is computed from a Gaussian approximation 
or the legendre transform of the beam maps or a more complex modelling
if the beams are asymetric, such as in~\cite{asymfast}.
\item The filtering transfer function is computed using a signal only 
Monte-Carlo simulation (it should include the pre-processing applied
to the time streams). Fake CMB sky are passed through the instrumental
and analysis process producing maps and pseudo power spectra. The
transfer function is basically computed as the ratio of the input
model to the recovered ensemble average. An important point at this
step is to check that the transfer function is independent of the
model assumed for the simulation. Let's also remark that using a
transfer function that depends only on $\ell$ is a bit daring as the
filtering is done in the scan direction which, in general corresponds
to a particular direction on the sky. This approximation however seems
to work well and has been successfully applied to
Boomerang~\cite{boom} and Archeops~\cite{arch_cell}).
\item The noise power spectrum is computed from noise only simulations
passing again through the instrumental and analysis process to produce
noise only maps and pseudo power spectra. The noise power spectrum is
estimated from the ensemble average of the various realisations.
\item Error bars are computed in a frequentist way by producing signal+noise 
simulations and analysing them as the real data. This allows to
reconstruct the full likelihood shape for each power spectrum bin and
the bin-bin covariance matrix.
\end{itemize}
Such an approach based on simulations has the advantage of being fast:
each realisation basically scales as $\propto N_t\log N_t$ for the
noise simulation and map-making (if filtering + coaddition is used)
and $\propto N_p^{3/2}$ for the CMB sky simulation and pseudo power
spectrum computation. An important advantage of such a method is the
possibility to include in the simulation systematic effects (beam,
pointing, atmosphere, ...) that would not be easily accountable for in
a maximum likelihood approach. 

\subsection{Cross-power spectra}
When several photometric channels are available from the experimental
setup, it is possible to compute cross-power spectra between the
channel rather than power spectra of individual channel or of the
average of all channels. This has the advantage of suppressing the
noise power spectrum (but not its variance of course) that is not
correlated between channels and leaving the sky signal unchanged. The
cross-power spectrum of channels $i$ and $j$ is defined as:
\begin{equation}
C_\ell^{i,j}=\frac{1}{2\ell+1}\sum_{m=-\ell}^{\ell}a_{\ell m}^{i}a_{\ell m}^{j \star}
\end{equation}
The cross-power spectrum method can easily be associated with the
frequentist approach simplifying significantly it implementation as
one the most difficult part, the noise estimation, is now less crucial
as noise disapears and cannot bias the power spectrum estimation. This
has been successfuly applied in the WMAP analysis~\cite{wmap_cell}.

\subsection{Which power spectrum estimator should be used ?}
The maximum likelihood approach is undoubtly the best method to use if
possible, but its CPU/RAM/Disk requirements are such that in practice,
with modern experiments, it is very difficult to implement. It should
however be considered to check the results on data subsets small
enough to make it possible. The frequentist approaches are much faster
and provide comparable precision in terms of error bars and permit to
account for systematic effects in a simple manner. The tricky part is
however to estimate the noise statistical properties precisely
enough. The same difficulty exists however in the maximum likelihood
approach where the noise covariance matrix has to be known
precisely. It is generally directly computed in the map-making process
from the time correlation function, thus displacing the difficulty
elsewhere. In any case, the noise model has to be unbiased as the
final power spectrum is essentially the subtraction between the pseudo
power spectrum of the map and the noise power spectrum. Estimating the
noise properties is a complex problem mainly due to signal
contamination and pixellisation effects. A general method for
estimating the noise in CMB experiment is proposed in~\cite{amblard}
and was successfuly applied for the Archeops
analysis~\cite{arch_cell}. When multiple channels are available, the
frequentist approach applied on cross-power spectra is certainly the
simplest and most powerful power spectrum estimation technique
available today as it reduces the importance of the difficult noise
estimation process. 
We can also mention the 
hierarchical decomposition~\cite{hierarch} that achieves an exact power 
spectrum estimation to submaps at various resolutions, and then optimally 
combine them. 

\section{Conslusions}
We have shown techniques designed to make maps from CMB data and to
extract power spectra from them. In both cases, the brute force,
maximum likelihood approach is the most correct, but generally hard to
implement in practice. Alternative approaches, iterative or relying on
Monte-Carlo simulations provide similar precision with smaller
computer requirements. In all cases, a lot of work has to be done before : first by designing the instrument
correctly and afterwards by cleaning the data, flagging bad samples
and ending up with a dataset that match the minimum requirement of all
the methods described in this review : stationarity and Gaussianity.

\Acknowledgements{The Author wants to thank the Archeops collaboration for 
its stimulating atmosphere, A.~Amblard and P.~Filliatre for reading
carefully the manuscript and providing useful inputs.}

%
\end{document}